\pdfoutput=1
\documentstyle[bbm,amsfonts,epsfig,12pt]{article}

\newcommand {\eqref} [1] {(\ref {#1})}
\newcommand {\slsh} [1] {\not{\hbox{\kern-2pt${#1}$}}}

\newcommand {\beq} {\begin{equation}}
\newcommand {\eeq} {\end{equation}}
  \newcommand {\ber}{\begin{eqnarray*}}
  \newcommand {\eer} {\end{eqnarray*}}
\newcommand {\beqn}{\begin{eqnarray}}
  \newcommand {\eeqn} {\end{eqnarray}}

\newcommand{\Dslash}{\,{\raise.15ex\hbox{/}\mkern-12mu D}}

\newcommand{\gsim}{\lower.7ex\hbox{$
\;\stackrel{\textstyle>}{\sim}\;$}}
\newcommand{\lsim}{\lower.7ex\hbox{$
\;\stackrel{\textstyle<}{\sim}\;$}}

\begin{document}
\begin{titlepage}
\begin{flushright}{UMN-TH-2720/08\,,  \,\,\, FTPI-MINN-08/38\,,  \,\, SLAC-PUB-13434}
\end{flushright}
\vskip 0.5cm

\centerline{{\Large \bf   Confinement in Yang--Mills:}}

\vspace{2mm}

\centerline{{\Large \bf   Elements of a Big Picture}}

\vskip 1cm
\centerline{\large  M. Shifman${}^{a,b}$  and Mithat \"{U}nsal ${}^{c,d}$}

\vskip 0.3cm

\centerline{${}^a$   \it William I. Fine Theoretical Physics Institute,}
\centerline{\it University of Minnesota, Minneapolis, MN 55455, USA}
\begin{center}
$^b$ {\it Laboratoire de Physique Th\'eorique\footnote{Unit\'e Mixte
de Recherche du CNRS,  (UMR 8627).}
Universit\'e de Paris-Sud XI\\
B\^atiment 210, F-91405 Orsay C\'edex, FRANCE}
\end{center}
\vskip 0.2cm
\centerline{${}^c$ \it SLAC, Stanford University, Menlo Park, CA 94025, USA}
\vskip 0.1cm
\centerline{${}^d$ \it  Physics Department, Stanford University, Stanford, CA,94305, USA }

\vskip 1cm

\begin{abstract}
This is a combined and slightly expanded version of talks delivered at
14th International QCD Conference ``QCD 08,"
7-12th July 2008, Montpellier, France,
the International Conference ``Quark Confinement 
and the Hadron Spectrum,"
Mainz, Germany, September 1-6, 2008 (Confinement 08), and 
the International Conference ``Approaches to 
Quantum Chromodynamics,"
Oberw\"{o}lz, Austria, September 7-13, 2008.

\end{abstract}

\end{titlepage}


\newpage

\section{Introduction}

QCD is a   non-Abelian gauge theory of strong interactions. It is extremely rich and 
 describes a very wide range of natural phenomena, e.g.:

$\bullet$ all of nuclear physics;

$\bullet$ Regge behavior and Regge trajectories;

$\bullet$ strongly coupled quark-gluon plasma; high-$T$ and high-density phenomena; neutron stars;

$\bullet$ richness of the hadronic world (chiral phenomena,
light and heavy quarkonia, glueballs and exotics,
exclusive and inclusive processes, interplay between strong and 
weak interactions, and many other issues).

At short distances QCD is weakly coupled, allowing high precision 
perturbative (multi-loop, multi-leg) calculations, 
however, analytical computations at all energy and momentum 
scales seem unlikely due to strong coupling nature  at large distances.
Let us ask ourselves: what do we want from this theory? Is it reasonable to
expect  high-precision predictions  for the low energy observables 
like in QED? Can we (an should we) compute  hadronic masses, matrix elements or 
proton magnetic moment up to, say, five digits? 

The answer to the last question, as well as other similar questions, seems to be negative, at least by analytical means.  Thankfully, at this front, 
exceedingly more precise and reliable numerical computations 
come from the lattice gauge theory. 
Lattice is the first-principles numerical 
framework for strongly coupled gauge theories and QCD.  
But, at the same time, it is a black box, ``an experiment."\footnote{ Lattices provide
non-perturbative numerical data; analytical theorists must engineer certain regimes for both 
QCD  and lattice theories  in which analytical results can be confronted with the lattice simulations. }

What we really need is a qualitative understanding of non-Abelian gauge dynamics
in various environments and various settings, an overall ``Big Picture."  

The overall picture that emerged in the last three decades -- and, especially, in the last ten years
or so -- makes theorists (who are still active in this area) rather happy.  A wealth of semi-classical techniques were developed in the past.
More  recently, some 
novel techniques allowing one to study non-perturbative  gauge dynamics through ``smooth transitions"
were devised.  These techniques  work even for notoriously elusive chiral gauge theories (for which no practical lattice formulations  exist). 
They provide ample opportunities for comparison of our 
analytical understanding of non-perturbative  gauge dynamics 
with the lattice theory. 
    
In the Big Picture, there are still  some dark corners to be explored, which will be mentioned
later. First, let us highlight general features of confinement and the  
elements which can be considered as well-established.

\section{Color confinement: Generalities}

The most salient feature of pure Yang-Mills theory  is linear confinement.
 If one takes a heavy probe quark and
an antiquark separated by a distance, the force between them does not fall off with distance,
while the potential energy grows linearly.

Are there physical phenomena in which interaction energy between two interacting bodies grows with distance at large distances? What is the underlying mechanism?

The answer to this question is positive. The phenomenon
 was predicted by Abrikosov 
\cite{ANO} in
the superconductors of the second type which, in turn, were invented by Abrikosov
\cite{ANL}
and discovered experimentally in the 1960s. The corresponding set up is shown in Fig.~1.
In the middle of this figure we see a superconducting sample, with two very long magnets attached to it.
The superconducting medium does not tolerate the magnetic field. On the other hand, the flux of the magnetic field
must be conserved. Therefore, the magnetic field lines 
emanating from the $S$ pole of one magnet
find their way to the $N$ pole of another magnet, through 
the medium, by virtue of a flux tube formation.
Inside the flux tube the Cooper pair condensate vanishes 
and superconductivity 
is ruined. The flux tube has a fixed tension, implying a constant force between the
magnetic poles as long as they are inside the superconducting sample. 
The phenomenon described above is sometimes referred to as the Meissner effect.

\begin{figure}[h]
  \begin{center}
  \includegraphics[width=5in]{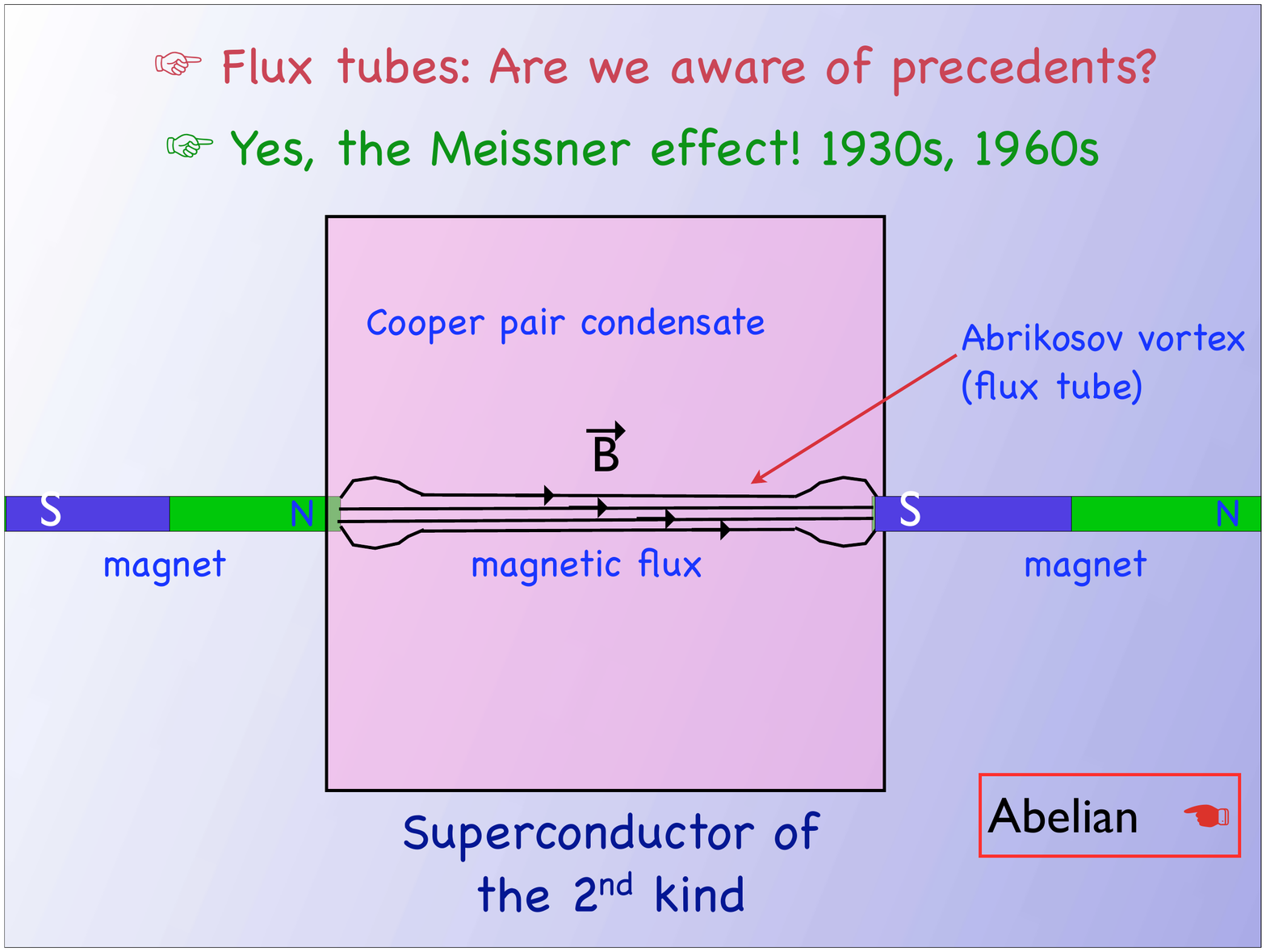}
  \caption
    {\small
The Meissner effect in QED
}
   \end{center}
\label{D1}
\end{figure}

Of course, the Meissner effect of the Abrikosov  type
occurs in the Abelian theory, QED. The flux tube that forms in this case is Abelian.
In Yang--Mills theories we are interested in its  non-Abelian analog.
Moreover, while in the Abrikosov case the flux tube is that of the magnetic field,
in QCD and QCD-like theories the confined objects are the quarks;
therefore, the flux tubes must be ``chromoelectric" rather than chromomagnetic.
In the mid-1970s Nambu,   't Hooft, and  Mandelstam (independently) 
put forward an idea \cite{NTM}
of a ``dual Meissner effect" as the underlying mechanism for color confinement  
(Fig.~2).\footnote{While Nambu and Mandelstam's publications
are easily accessible, it is hard to find the EPS Conference Proceedings
in which 't Hooft presented his vision. Therefore, the corresponding passage from his talk
is worth quoting: ``...[monopoles] turn to develop a non-zero vacuum expectation value.
Since they carry color-magnetic charges, the vacuum will behave like a superconductor
for color-magnetic charges. What does that mean?
Remember that in ordinary electric superconductors, magnetic charges are confined by magnetic vortex lines ... We now have the opposite: it is the color charges that are confined by electric flux tubes."
}
Within the framework of this idea, in QCD-like theories ``monopoles" condense
leading to formation of ``non-Abelian flux tubes" between the probe quarks.
At this time the  Nambu--'t Hooft--Mandelstam paradigm was not even a physical 
scenario, rather a
dream, since people had no clue as to
the main building blocks such as non-Abelian chromoelectric
flux tubes. After the Nambu--'t Hooft--Mandelstam
conjecture many works had been published on this subject,
with very little advancement, if at all.

\begin{figure}[h]
  \begin{center}
  \includegraphics[width=5in]{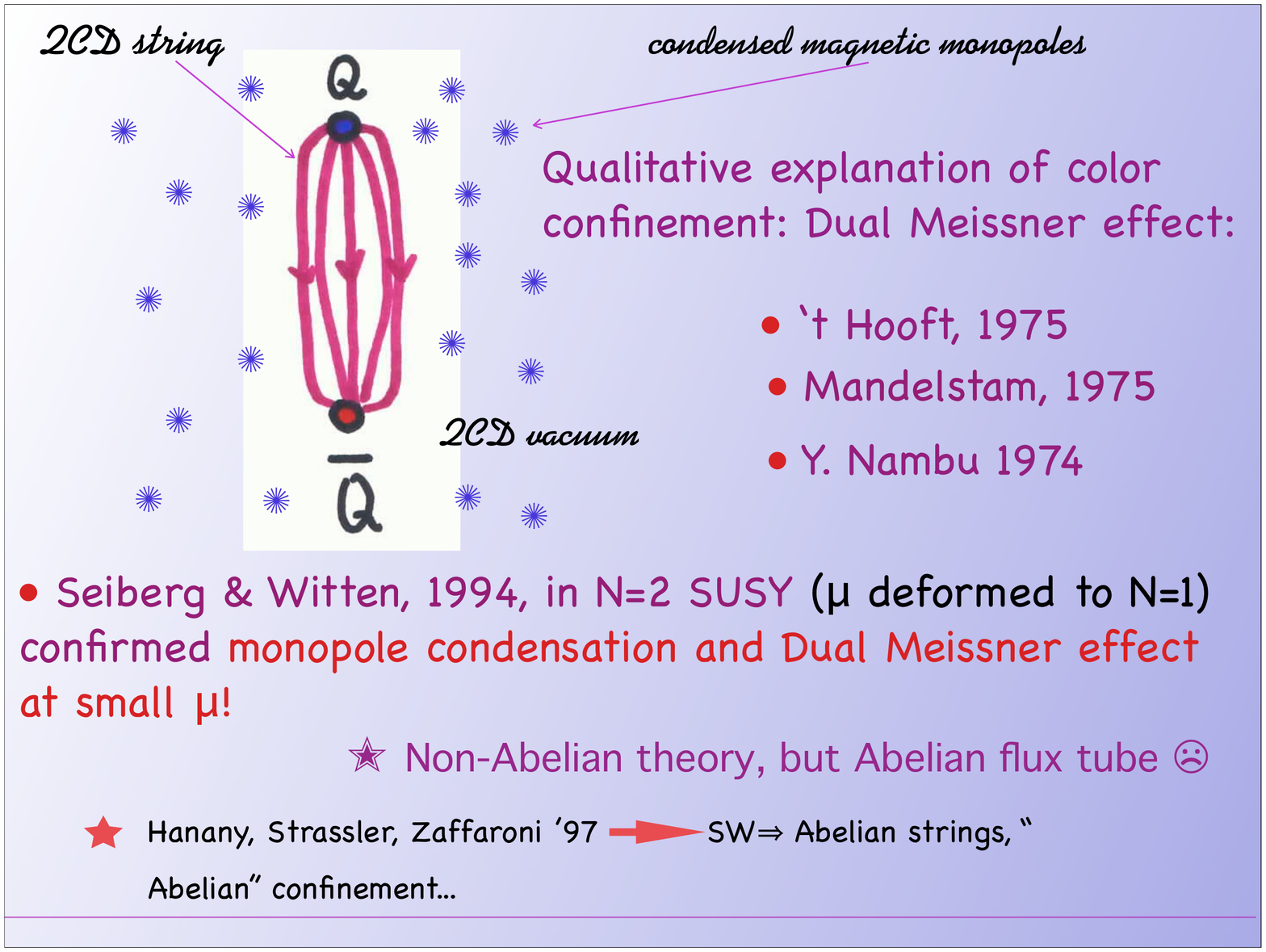}
  \caption
    {\small
The Meissner effect in QED
}
   \end{center}
\label{D2}
\end{figure}

A decisive breakthrough came in 1994, with the Seiberg--Witten solution of
${\mathcal N}=2$ super-Yang--Mills 
\cite{SW}
slightly deformed by a $\mu \, {\rm Tr}\Phi ^2$ term
in the superpotential. The deformation term breaks ${\mathcal N}=2$ down to ${\mathcal N}=1$.
At $\mu =0$ the theory has a moduli space parametrized by
${\rm Tr}\Phi^2$
(we assume for simplicity that the gauge symmetry of the action is SU(2)).
On the moduli space,
SU(2)$_{\rm gauge}$ is spontaneously broken down to U(1). Therefore, the theory possesses 
the 't Hooft--Polyakov monopoles \cite{TP}.
Two points on the moduli space were detected in \cite{SW} (the so-called monopole and dyon points)
in which the monopoles (dyons) become massless. 
In these points  the scale of the gauge symmetry breaking 
\beq
{\rm SU}(2)\to {\rm U}(1)
\label{one}
\eeq
 is determined by the dynamical parameter
$\Lambda$ of the microscopic theory. While the neutral gauge boson (photon) remains massless,
others acquire masses of the order of $\Lambda$ and can be integrated out.
In the low-energy limit near the monopole and dyon points
one deals with electrodynamics of massless monopoles. One can formulate an effective macroscopic
low-energy theory. This is a U(1) gauge  theory 
in which the charged matter fields $M$, $\tilde M$ are those of monopoles while the
gauge field is dual with respect to the photon of the microscopic theory.
The superpotential has the form ${\mathcal W} = {\mathcal A} M\tilde M$,
where ${\mathcal A}$ is  the ${\mathcal N}=2$ superpartner of the dual photon/photino fields.

Now, if one switches on $\mu \neq 0$ (and $|\mu | \ll \Lambda$),
the only change in the macroscopic theory is the emergence of the extra $m^2{\mathcal A} $
term in the superpotential. The dual mass parameter $m^2\sim \mu\Lambda$.
The $m^2{\mathcal A} $ term triggers the monopole condensation,
$\langle M\rangle = \langle \tilde M\rangle =m$, which implies, in turn,
that the dual U(1) symmetry is spontaneously broken, and the dual photon acquires a mass
$\sim m$. As a consequence, Abrikosov flux tubes
are formed. Viewed inside the dual theory, they carry fluxes
of the magnetic field. With regards to the original microscopic theory
these are the electric field fluxes.

Thus, Seiberg and Witten demonstrated, for the first time ever, the existence of the dual Meissner
effect in a judiciously chosen non-Abelian gauge field theory. 
If one ``injects" a probe (very heavy) quark and antiquark in this theory,
a flux tube necessarily forms
between them
leading to linear confinement. In the leading order in $\mu$ this flux tube is BPS saturated.
Its tension $T$ is  proportional to $ \mu\Lambda$.

The flux tubes in the Seiberg--Witten solution were investigated in detail
in 1997 by Hanany, Strassler and Zaffaroni \cite{HSZ}.
These flux tubes are Abelian, and so is confinement caused by their formation.
What does that mean? At the scale of distances at which the flux tube is formed
(the inverse mass of the Higgsed U(1) photon) the gauge group that is operative
is Abelian. In the Seiberg--Witten analysis this is the dual U(1). 
The off-diagonal (charged) gauge bosons are very heavy in this scale
and play no direct role in the flux tube formation and 
confinement that ensues. Naturally, the spectrum of composite objects in this case
turns out to be richer than that in QCD and similar theories with non-Abelian
confinement.\footnote{Anticipating further discussions let us
note that by  non-Abelian
confinement we mean such dynamical regime in which at distances
of the flux tube formation all gauge bosons are equally important.}
It includes not only color singlets, but, in addition, a variety of states
which are  U(1) neutral rather than SU(2) neutral. Moreover, the string topological stability is based
on $\pi_1 ({\rm U}(1)) = Z$. Therefore, $N$ strings  
do not annihilate as they should in QCD-like theories. 

The double-stage symmetry breaking pattern, with (\ref{one}) occurring at a high scale
while the dual U(1) $\to$ nothing 
 at a much lower scale, disappears as we let $\mu$ approach $\Lambda$. 
Eventually we could send this parameter to infinity.
In the limit $\mu\to\infty$ we would recover an ${\mathcal N} =1$ rather than 
${\mathcal N} =2$ theory
in which all non-Abelian gauge degrees of freedom presumably 
take part in the string formation,
and are operative at the scale at which the strings are formed. The strings that
occur (if they occur) may be called non-Abelian. 

The Seiberg--Witten solution {\em per se}
is applicable
only at $|\mu | \ll \Lambda$. We have no quantitative (or even semi-quantitative)
description of the non-Abelian confinement emerging  at large $|\mu |$,
when the U(1) subgroup of SU(2) is no longer singled out.
It is generally believed, however, that the transition from the Abelian to  non-Abelian regime
is smooth. It is argued that  the strings occurring in the 
Seiberg--Witten solution belong to the same universality class as those in QCD-like theories.
We will return to the issue of universality classes and non-Abelian flux tubes later, in 
Sect.~\ref{pfnaf}.
Our immediate task is to develop a similar strategy for non-supersymmetric  
QCD-like gauge theories which  (by definition)
do not possess elementary scalars in the microscopic Lagrangian. 

\section{QCD-like theories on a cylinder; double-trace deformation}

 In the Seiberg--Witten case the deformation parameter
 which governs the transition from Abelian to non-Abelian confinement
 is $\mu$, the parameter of the ${\mathcal N} =2$ supersymmetry breaking.
 In the non-supersymmetric case we have to invent another parameter
 which would play the same role. 
 An appropriate set-up was suggested by  Shifman and  \"Unsal \cite{ShifUns}.
 Instead of considering QCD and QCD-like cousins on $R_4$
 one can compactify (say, a spatial) dimension, formulating the theory on a cylinder 
 $R_3\times S_1$ (Fig.~3). The radius of the compact dimension $r(S_1)$
 is the parameter regulating dynamical regimes of the theory.

\begin{figure}[h]
  \begin{center}
  \includegraphics[width=5in]{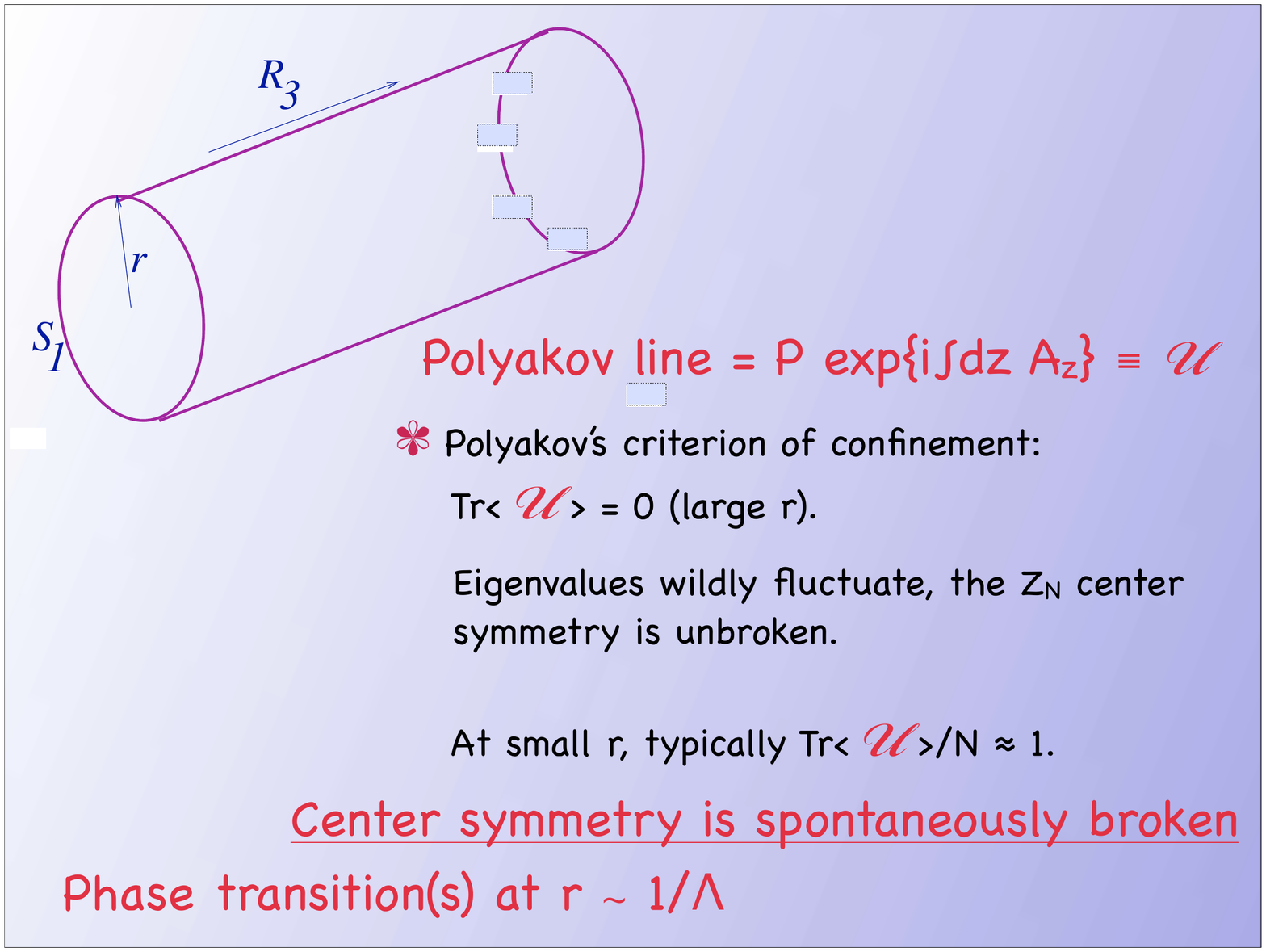}
  \caption
    {\small
QCD-like theories on a cylinder.
}
   \end{center}
\label{D3}
\end{figure}

Of course, in the past such a set-up was considered many times,
for various purposes,
with a number of distinct boundary conditions, with the temporal or spatial compactification.
It did not bring people closer to the goal we have in mind
today. The point is that the small-$r(S_1)$ domain was always separated from
the decompactification
limit of large $r(S_1)$ by one or more phase transitions (Fig.~4).

\begin{figure}[h]
  \begin{center}
  \includegraphics[width=5in]{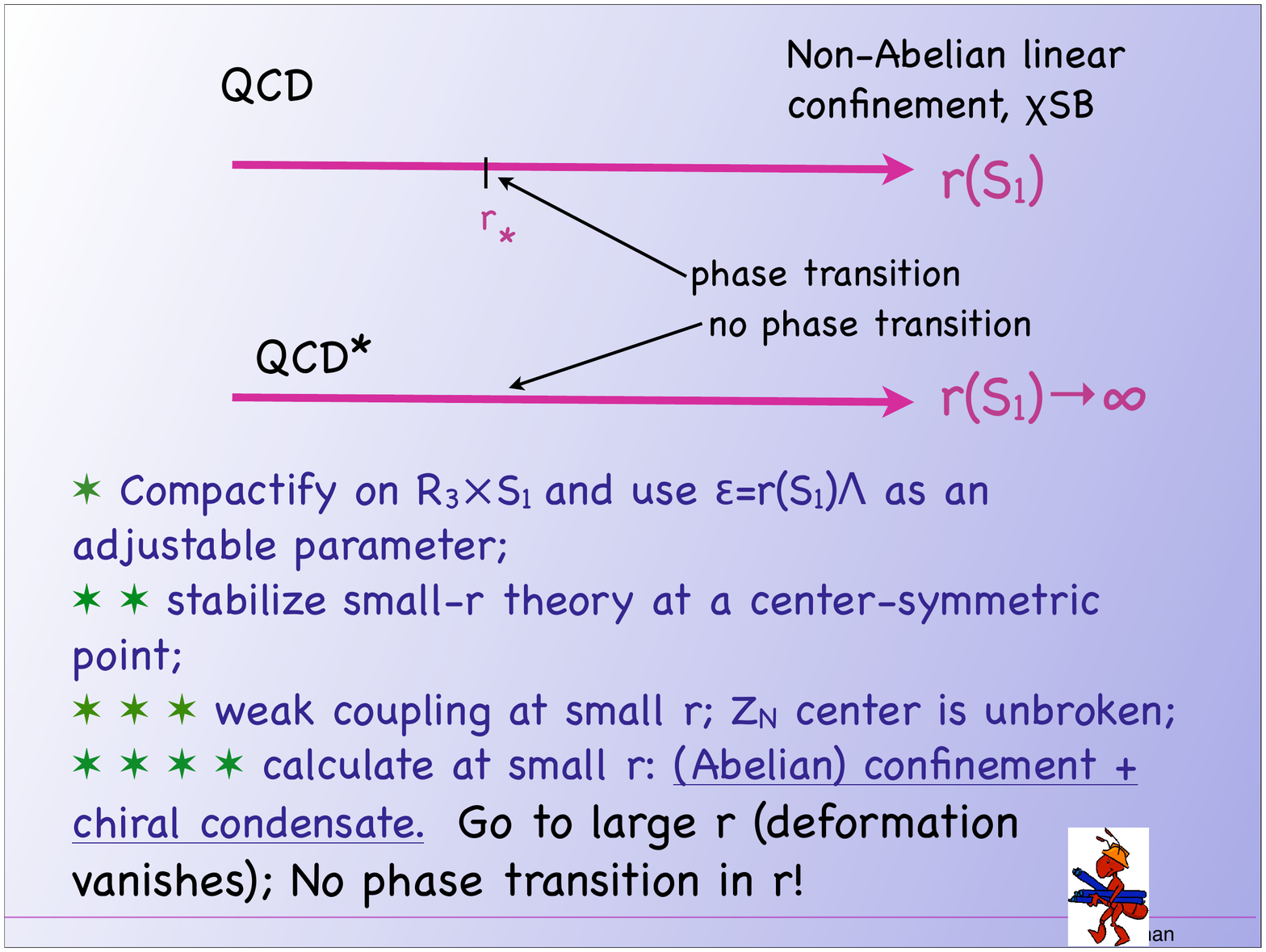}
  \caption
    {\small
Proceeding from small- to large-$r(S_1)$ one usually experiences
a confinement-deconfinement (and, possibly, chiral symmetry restoration)
phase transition. To make the $r(S_1)$ evolution smooth we add double trace deformations.
}
   \end{center}
\label{D4}
\end{figure}

In particular, in pure Yang--Mills theory a confinement-deconfinement phase transition
was identified at $r(S_1)= r_*  \sim\Lambda^{-1}$. Moreover, with massless quarks included,
the large-$r(S_1)$ domain, with the spontaneously broken chiral symmetry,
is separated from the small-$r(S_1)$ domain, with a restored chiral symmetry,
by a chiral phase transition, at (approximately) the same
value of  $r(S_1)\sim r_*$. The only exception is  ${\mathcal N} =1$ super-Yang--Mills endowed 
with the periodic boundary condition. 

The reason of the unwanted phase transition(s)
is the breaking of the center symmetry in the conventional (thermal) set-up at small $r(S_1)$.
The famous Polyakov criterion \cite{Polyak,Sussk} tells us then that the theory is in the deconfinement phase. Moreover, 
the vacuum structure is such that even  at small $r(S_1)$
the theory is governed by the strong-coupling regime which precludes analytic control.

The physical picture in the confinement phase is as follows. Assume that the  compactified dimension is $z$.
The Polyakov line (sometimes called the Polyakov loop)
 is defined as a path-ordered holonomy of the Wilson line
in the compactified dimension,
\beq
{\cal U} = P\exp\left\{i\int_0^L a_z dz \right\} \equiv V U V^\dagger
\label{onem}
\eeq
where $L$ is the size of the compact dimension while
$V$ is a matrix diagonalizing ${\cal U}$,
\beq
U = {\rm diag}\{ v_1, v_2, ..., v_N\} \,.
\label{twom}
\eeq
According to Polyakov, non-Abelian confinement implies that the eigenvalues 
$v_i$ are randomized:  the phases of $v_i$ wildly fluctuate over the entire
interval $[0,2\pi]$ so that 
\beq
\langle {\rm Tr} U \rangle =0\,.
\label{threem}
\eeq
The vanishing of $\langle {\rm Tr}\, U \rangle$ implies that the center symmetry is unbroken.

The double-trace deformations suggested in this context in
\cite{ShifUns} eliminate both unwanted features at once. 
A general design is as follows.
On  $R_3\times S_1  $ at small $r(S_1)$ one can deform the original  theory (for the time 
being we mean pure Yang--Mills or one-flavor QCD) by a  double-trace operator $P[U_J({\bf x}) ]$
where 
 \begin{equation}
P[U({\bf x}) ]= \frac{2}{\pi^2 L^4} \, \sum_{n=1}^{\left[\frac{N}{2}\right]}  d_n        
 \left|
\, {\rm Tr}\, U^n ({\bf x} )\right|^2 \,.
\label{fourma}
\end{equation}
Here
 $d_n$ are numerical parameters of order one, and ${\left[...\right]} $ denotes the integer part of 
the argument in the brackets. 
The deformed action is 
\begin{equation}
S^{*} = S + \int_{R_3 \times S_1} P[U({\bf x}) ]
\end{equation} 
We label the deformed theories with an asterisk (e.g. YM*). 
For judiciously chosen  $d_n$, the center  symmetry remains unbroken in the vacuum.
This is due to the fact that the gauge symmetry SU$(N)$ spontaneously breaks
\beq
{\rm SU}(N)\to {\rm U}(1)^{N-1},
\label{higgsed}
\eeq
with the pattern of the eigenvalues of $U$ depicted in Fig. 5.  This means, the 
long-distance dynamics is effectively Abelian.  

 At strong coupling  confinement takes place because the eigenvalues of 
$U$ wildly fluctuate implying  vanishing of the Polyakov line.  
These wild fluctuations are in one-to-one correspondence with
formation of confining flux tubes. At weak coupling the Polyakov line vanishes for a different reason, 
\begin{figure}[h]
  \begin{center}
  \includegraphics[width=5in]{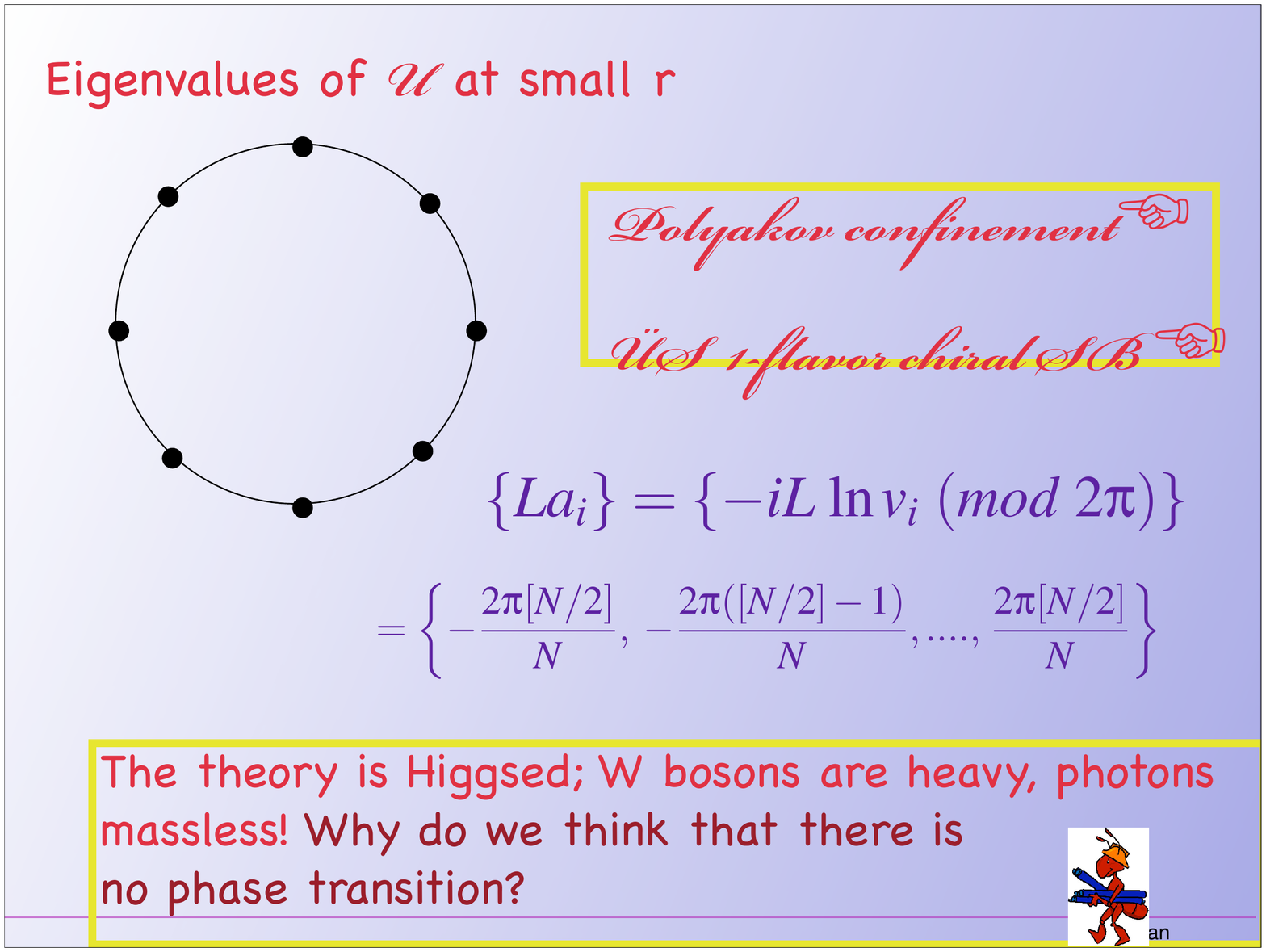}
  \caption
    {\small
Center symmetry stabilization by double trace deformations.
}
   \end{center}
\label{D5}
\end{figure}
but this is still a confinement.

\section{Linear confinement at small $r(S_1)$}

Consider the deformed Yang--Mills theory.  
The SU$(N)$ gauge group is Higgsed down  to an Abelian
${\rm U}(1)^{N-1}$ Cartan subgroup at the scale of compactification (remember, $r(S_1)$
is small for the time being). 
 At low energies, keeping only
the zero modes, we end up with $N-1$ ``photons" in three dimensions.
Perturbatively, they are massless. Correspondingly,
there is no linear confinement to  all orders in perturbation theory. 

The picture drastically changes at the non-perturbative level.
In  Yang--Mills-(adjoint)Higgs system this is known from 1977, when Polyakov demonstrated that
three-dimensional instantons generate a mass
for the dual photons \cite{Pol}.\footnote{Some elements of this construction were presented 
by Polyakov in his 1975 paper \cite{Polyakov:1975rs}, see the concluding paragraphs.
}
  Technically, these instantons look exactly as monopoles in 
four dimensions; hence we will refer to them as instanton-monopoles. 

It is obvious that in 2+1 dimensions the photon field
has only one physical (transverse) polarization.
This means that the photon field must have a dual description
in terms of one scalar field $\varphi$, namely \cite{Pol}
\beq
F_{\mu\nu}
 = k \, \varepsilon_{\mu\nu\rho}\left( \partial^\rho\,\varphi\right)\,,
 \label{14.8}
\eeq
where $k=g^2_{3D}/4\pi$ and $g^2_{3D} = g^2_{4D} \,L^{-1}$. The circumference of $S_1$
is $L=2\pi r(S_1)$.
Equation (\ref{14.8}) defines the field $\varphi$
in terms of $F_{\mu\nu}$ in a non-local way.
At the same time, $F_{\mu\nu}$ is related to $\varphi$ locally.

Consider a probe heavy charge $\pm 1/2$
 at the origin.
The electric field induced by the probe particle is radially oriented.
A brief inspection of Eq.~(\ref{14.8}) shows the the radial orientation of $\vec E$
requires the scalar function $\varphi$ to be $r$-independent.
Moreover, it should depend on the polar angle $\alpha$ as
\beq
\varphi = \alpha\,.
 \label{14.10}
\eeq
Needless to say, Eq.~(\ref{14.10}) implies that the scalar
field $\varphi$ is compact and defined mod $2\pi$. The points
\beq
\varphi,\quad \varphi\pm 2\pi,\quad  \varphi\pm 4\pi, ...
 \label{14.11}
\eeq
are identified. Thus, Polyakov's observation that in 2+1 dimensions
the photon field is dual to a real scalar field
needs an additional specification: the real scalar field at hand is compact;
it is defined on a circle of circumference $2\pi$.\,\footnote{The non-compact (free) 
Maxwell theory is dual to a free scalar theory with a continuous shift symmetry. Compact 
electrodynamics has instantons; this explicitly breaks the continuous shift symmetry 
down to a $2 \pi$ shift symmetry. The latter can no longer protect masslessness of the dual scalar.}
The
minimal electric charge in the original formulation is equal to the  minimal vortex in the dual formulation, and vice versa.

As was noted by Polyakov, the instanton-monopole contribution
generates a mass term for the field $\varphi$ , 
\beq
{\cal L}_{\rm dual}
= \frac{\kappa^2}{2}\,\left(\partial_\mu\varphi\right) \left(\partial^\mu\varphi\right) + \mu^3\, \cos  \varphi  \,,
\label{14.32}
\eeq
where $\kappa =g_{3D}/4\pi$ and 
\beq
\mu^3 \,\propto \, e^{-S_{\rm inst}}
\eeq
where
\beq
S_{\rm inst}= \frac{4\pi m_W}{g^2_{3D}}\,.
\eeq
Hence,  $\mu^3$ is exponentially small at small $r(S_1)$.
 The dual photon mass 
$m_\varphi = \mu^{3/2}\,\kappa^{-1}$ is exponentially small too.

In the dual language it is quite obvious that if $m_\varphi \neq 0$,
there exist domain ``lines" in 1+2 dimensions, a.k.a. strings,
separating the (physically equivalent) domains
$\varphi =0$ and $\varphi =2\pi$ (or, more generally,  $\varphi_{\rm vac} =2\pi \, k  
$ and $\varphi_{\rm vac} = 2\pi (k+1)$
where $k$ is an arbitrary integer), see Fig.~6. The endpoints of the ``line"
are $\varphi$ vortices. The line thickness is $\ell\sim 1/m_\varphi$.
\begin{figure}[h]   
\epsfxsize=4cm
\centerline{\epsfbox{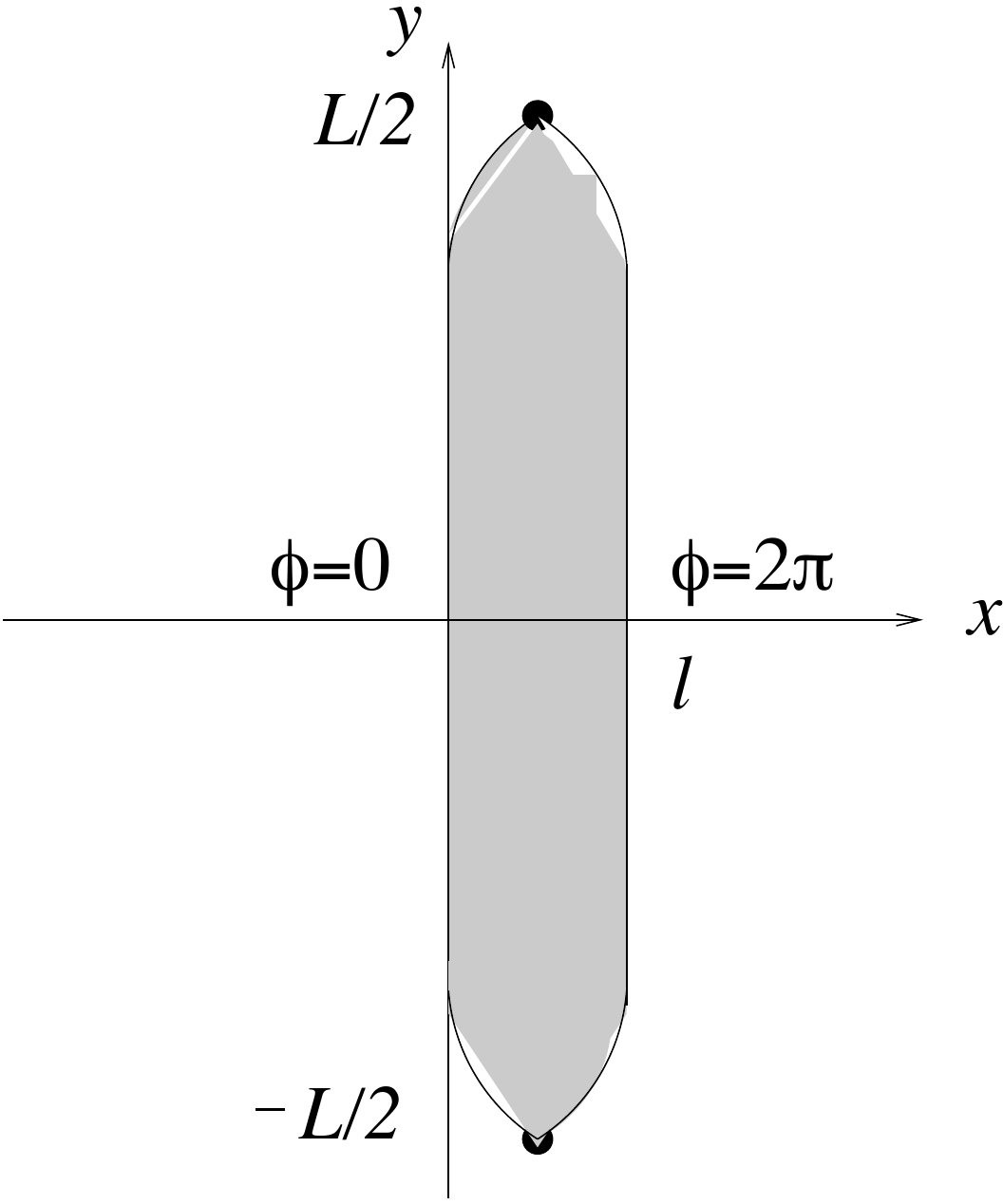}}
\caption{``Domain wall" in 2+1 dimensions. The  circles at the endpoints 
represent probe charges $\pm 1/2$. These are vortex-anti-vortex pairs in the dual formulation.
The picture assumes that $L\gg\ell$. The transitional domain
which is a domain line and a string simultaneously, is shaded.}
\label{f14three}
\end{figure}

The Polyakov string tension 
is
\beq
T = 8\mu^{3/2}\,\kappa = 8 \, m_{\varphi}\,\kappa^2 = 2\, k\, m_{\varphi}\,.
\label{14.36}
\eeq
Note that this tension is much larger than $\ell^{-2}$.

For this string to develop between two probe charges the
distance $L$ between the charges must be $L\gg \ell$.
At distances $\lsim \ell$ each charge is surrounded by essentially
a (two-dimensional) Coulomb field, with the force lines spreading homogeneously in all directions. At distances $\sim \ell$
the ``flux tube" starts forming. If
the probe charges have opposite signs, and  $L\gg \ell$, 
they are connected by the ``flux tube," and 
the energy of
the configuration grows as $T\, L$. At   huge distances $L\gg \ell$
the linear confinement sets in.

The   YM*  theory at  small $r(S_1)$ can be viewed as a very similar three-dimensional
 YM--Higgs system, in which a compact adjoint Higgs field -- 
 the holonomy -- replaces the non-compact field of the Polyakov model. 
 This ``compactness"  introduces  an extra topological excitation (see e.g. \cite{Davies}). 
 Some technicalities of the analysis change; but, in essence,
 we have the same dynamical system. However, introduction of massless quarks  
 drastically changes the picture.

\section{Massless quarks}

Let us speak for definiteness of a single massless quark in the SU($N$) gauge theory:
either in the fundamental representation (for QCD) or in the two-index 
representations (for QCD-like theories). If such a massless quark is
introduced in the Polyakov three-dimensional model {\em per se}
 the model ceases to be confining! This is due to the fact that the instanton-monopoles
 have zero fermion modes and no longer generate the dual photon mass.
Rather,
 they give rise to a chiral condensate which may  break a discrete 
chiral symmetry.

However, in QCD$^*$ on $R_3\times S_1$ the situation is different.
In this theory there are other, composite,  topological excitations (different from the 
instanton-monopoles)
which carry a net magnetic charge, but no topological charge and, hence, no fermion zero modes.
Such excitations do generate a potential and produce a mass term for dual photons,
implying linear confinement. 

In this aspect there is a crucial difference between
the Polyakov construction which is genuinely three-dimensional
and the $R_3\times S_1$ reduction of QCD and QCD-like theories. For example, 
in the compactified four-dimensional SU(2) Yang--Mills theory
there are two types of monopoles: one is the standard 
't Hooft--Polyakov (tHP) monopole; another can be called
the Kaluza--Klein (KK) monopole \cite{Davies}; its existence  is due to the fact that
$\pi_1 (S_1) = Z$.  In QCD-like theories with the
two-index representation fermions, the bound state of the tHP monopole and the KK antimonopole
carries no topological charge; hence, no fermion zero modes.
It does carry a magnetic charge and, therefore, generates a mass term for the dual photon.
The above pair was shown to be  stable \cite{Unsal:2007jx};  it was termed
a  magnetic bion. In SU$(N)$ theories the set of bions 
responsible for nonperturbative physics on $R_3\times S_1$
is quite varied;
they contribute to the dual photon effective potential (i.e. to the dual photon mass squared)
at the level $\exp (-2S_{\rm inst})$. 

Summarizing the situation in one-flavor QCD$^*$, the
instanton-monopoles generate the chiral condensate at the level
$\exp (-S_{\rm inst})$.  Bions are responsible for the dual photon masses
which lead, in turn, to (Abelian) linear confinement through domain lines
\cite{ShifUns}.  Four-dimensional instantons produce effects
which are suppressed at the level $\exp (-NS_{\rm inst})$. 

The  multiflavor QCD*  theories exhibit an interplay of different dynamics; 
it is currently under consideration.

\section{Transition to large $r(S_1)$}

Much in the same way as passing from small to large $\mu$ in the Seiberg--Witten
constructions is believed to take us smoothly from the Abelian confinement to the non-Abelian one, 
we would like to argue that
the transition to large $r(S_1)$ does the same job  in YM$^*$ and one-flavor QCD$^*$, and  is smooth too.

Looking from four dimensions we see the following order parameters:
the Polyakov line (confinement/deconfinement) and the
chiral condensate. The first vanishes both at small and large $r(S_1)$. Hence,
it gives no signal as to a possible confinement-deconfinement phase transition.
On the contrary, the second does not vanish both at small and large
$r(S_1)$, indicating spontaneous breaking of a discrete chiral symmetry.
Therefore, there is no reason to expect a chiral phase transition either
(for quarks belonging to higher representations).

At small $r(S_1)$ confinement is obviously Abelian. At large $r(S_1)$
we expect non-Abelian confinement, much in the same way as in the Seiberg--Witten problem 
at large $\mu$. There is no obvious obstacle to a smooth transition
from the Abelian confinement to that of the non-Abelian type.
From the four-dimensional standpoint we can indicate no appropriate order parameter.

Let us ask ourselves what quantitative parameters could guide us in the description
of evolution from the Abelian to non-Abelian confinement?
The source of quantitative information might be
 the effective theory on the string world sheet.
At small $r(S_1)$ the string is Abelian and the
effective Lagrangian is that describing transverse displacements of the string position.
If, in addition, there are hidden light degrees of freedom, which belong to this Lagrangian,
and make the vacuum structure on the string world sheet nontrivial,
then the phase transition on the string world sheet could separate
the small and large-$r(S_1)$ domains.

Of course, in two dimensions only discrete symmetries (e.g. $Z_N$) can be spontaneously broken.
There are no obvious reasons for the occurrence of $Z_N$ symmetric 
extra moduli on the string world sheet. That's why we believe
that the domain of the Abelian confinement at small $r(S_1)$
is not separated from the decompactification limit (non-Abelian confinement)
by phase transition(s). The center-symmetric construction we develop is
smooth.

\section{A few words on chiral theories}

Strongly coupled chiral Yang--Mills theories did not receive sufficient  attention that they deserve.
This is not because they are dynamically uninteresting (on the contrary,
they are interesting and may even be relevant for TeV scale physics) but, rather, 
because it is notoriously difficult to come up with a reliable quantitative framework for their analysis.   They are certainly the most mysterious ones among non-Abelian gauge theories. That we were
unable to say anything at all  about their non-perturbative dynamics was very disappointing. 

We will  split the chiral Yang--Mills theories in two classes.  A more traditional one,
which was to an extent addressed in the 1980s \cite{DimPes} is represented, e.g.
by the SU$(N)$ gauge theory with one two-index  antisymmetric  or symmetric (AS and S, for short)  Weyl fermion plus $N-4$ or  $N+4$ antifundamental  Weyl fermions,
$$ 
 \Big\{ \psi_{\{ab\}},  (N+4) \psi^a \Big\}\,\,\,{\rm   and  }\,\,\, \Big\{ \psi_{[ab]},  (N-4) \psi^a \Big\}.$$  

A  novel class of chiral (quiver)  gauge theories are  the $Z_K$ orbifold projections of SU$(KN)$
 supersymmetric gluodynamics with $K\geq 3$.  These are 
 perturbatively planar equivalent (PPE) to supersymmetric gluodynamics. Two other  types of new chiral theories are  SU$(N)$ gauge theories with 
 the following matter content:
  $$\Big\{ \psi_{\{ab\}}, \psi^{[ab]}, \,\, 8 \psi^a \Big\}\,\,\,{\rm   and  }\,\,\,\Big\{ \psi_{\{ab\}}, \psi_{[ab]}, \,\, 2N  \psi^a \Big\}.$$ 
  For the both classes, internal gauge anomalies cancel. For example, in the first case,  
 $d_{S} +  d_{\overline {AS}} + 8d_{\overline {F}}= N+4 - (N-4) - 8=0$, and these theories are consistent perturbatively.   These theories arise in the context of planar orientifold equivalence. 
 Note that these are essentially the two independent combinations of the chiral theories mentioned in the previous paragraph.

 Again, much in the same way as in the non-chiral theories,
we can compactify the  space-time onto $R_3 \times S_1  $ and perform double-trace deformations to stabilize the center-symmetric vacuum at small-$r(S_1)$. This allows one to
smoothly connect  small-$r(S_1)$ to large-$r(S_1)$ physics 
 where the double-trace deformations are switched off.
In this way bion-induced  linear confinement 
was analytically
demonstrated \cite{SUn}. A remarkably rich  pattern of the chiral symmetry realizations
which depends on the structure of the ring operators, a novel
class of topological excitations, was revealed in \cite{SUn} on the way!  
Perhaps, one of the most interesting aspect of these theories  is that the simplest  monopole-instanton operators drop out.  
Another surprising aspect of the  [SU(N)$]^K$ chiral quiver theories is that, 
although they  possess a $Z_{2N}$ axial chiral symmetry -- just like pure supersymmetric gluodynamics
-- this symmetry may or may not be broken depending on the relation between $N$ and $K$.  What does this imply for PPE? 

PPE extends to a  full  non-perturbative equivalence provided: (i)
 the orbifold projection symmetry is not spontaneously broken in the parent theory; and (ii)  the   $Z_K$  cyclic  symmetry is not spontaneously broken in the daughter quiver theory \cite{Kovtun:2003hr}.  
In our case with 
$K \geq 3$, the necessary and sufficient conditions are violated 
in the supersymmetric parent rather than in the non-supersymmetric daughter.  The $Z_K$ symmetry used in the orbifold projection is part of the discrete chiral symmetry which breaks down spontaneously  to $Z_2$. Consequently,  the non-perturbative equivalence fails in the chiral case $K \geq 3$,
while it is valid in the vector-like case $K=2$ \cite{ShifUns}.  
Our microscopic description of the chiral quiver theories also confirms this result. 

The chiral gauge dynamics is very different from vector-like or supersymmetric 
gluodynamics. We refer the reader to  \cite{SUn} for a very detailed description. 

\section{Prototypes for the non-Abelian strings}
\label{pfnaf}
Now we will briefly discuss a prototype
non-Abelian flux tube. Note that the flux it carries is that of the magnetic rather
than electric field. Dualization of this particular model is
not yet performed. The (non-supersymmetric) gauge theory supporting such strings 
\cite{GSY}
is at weak coupling, while
the theory on the string world sheet is at strong coupling (Fig. 7).
\begin{figure}[h]   
\epsfxsize=6cm
\centerline{\epsfbox{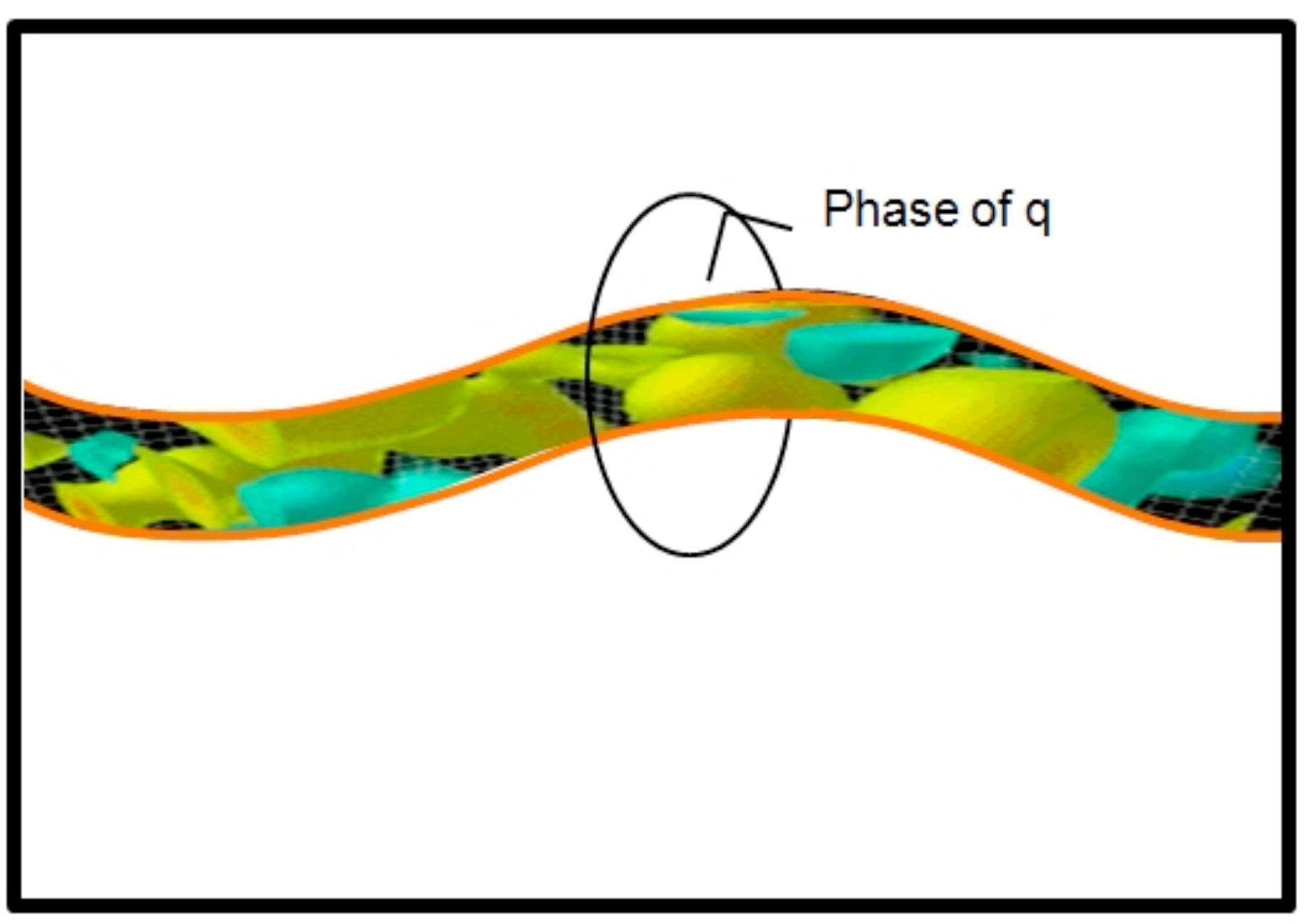}}
\caption{A prototype non-Abelian string in the model of Ref.~\cite{GSY}. This figure
is borrowed from \cite{Tong}.}
\label{f14three}
\end{figure}

A prototype model has the gauge group  
 U($N)$. Besides SU($N$) and U(1)
gauge bosons   
the model contains $N$ scalar fields charged with respect to
U(1) which form $N$ fundamental representations of SU($N$).
It is convenient to write these fields in the form of 
$N\times N$ matrix $\Phi =\{\varphi^{kA}\}$
where $k$ is the SU($N$) gauge index while $A$ is the flavor
index, 
\beq
\Phi =\left(
\begin{array}{cccc}
\varphi^{11} & \varphi^{12}& ... & \varphi^{1N}\\[2mm]
\varphi^{21} & \varphi^{22}& ... & \varphi^{2N}\\[2mm]
...&...&...&...\\[2mm]
\varphi^{N1} & \varphi^{N2}& ... & \varphi^{NN}
\end{array}
\right)\,.
\label{phima}
\eeq
The action of the model is
\beqn
S &=& \int {\rm d}^4x\left\{\frac1{4g_2^2}
\left(F^{a}_{\mu\nu}\right)^{2}
+ \frac1{4g_1^2}\left(F_{\mu\nu}\right)^{2}
+ {\rm Tr}\, (\nabla_\mu \Phi)^\dagger \,(\nabla^\mu \Phi )
 \right.
 \nonumber\\[3mm]
&+&
\left.
\frac{g^2_2}{2}\left[{\rm Tr}\,
\left(\Phi^\dagger T^a \Phi\right)\right]^2
 +
 \frac{g^2_1}{8}\left[ {\rm Tr}\,
\left( \Phi^\dagger \Phi \right)- N\xi \right]^2 \right\}
\label{redqed}
\eeqn
where $N$ stands for the number of colors and
the parameter $\xi$ which has dimension $m^2$
is assumed to be large which guarantees weak coupling.
The last 
term in the second line
forces $\Phi$ to develop a vacuum expectation value (VEV) while the 
last but one term
forces the VEV to be diagonal,
\beq
\Phi_{\rm vac} = \sqrt\xi\,{\rm diag}\, \{1,1,...,1\}\,.
\label{diagphi}
\eeq
Thus, the model is fully Higgsed in the bulk; there are no massless excitations.
A diagonal global SU($N$) survives, however,
namely,
\beq
{\rm U}(N)_{\rm gauge}\times {\rm SU}(N)_{\rm flavor}
\to {\rm SU}(N)_{\rm diag}\,.
\label{vacs}
\eeq
Thus, color-flavor locking takes place in the vacuum.

The string solution can be obtained by winding one of the elements of the field
matrix (\ref{diagphi}). Its topological stability is due to the fact
that 
\beq
\pi_1 \left({\rm SU}(N)\times {\rm U}(1)/ Z_N
\right)\neq 0\,.
\eeq

The string solution 
breaks the symmetry (\ref{vacs}) 
down\,\footnote{At $N=2$ the string solution breaks
SU(2) down to U(1).} to U(1)$\times$SU$(N-1)$ (at $N>2$).
This means that the world-sheet (two-dimensional) theory of 
the elementary string moduli
is the SU($N$)/(U(1)$\times$ SU($N-1$)) sigma model.
This is the famous $CP(N-1)$ model. It is strongly coupled in the infrared 
and develops its own dynamical 
scale $\Lambda$. 

This string is formed at the distances of the order of the inverse gauge boson
masses (which are all of the same order if we do not consider $N$ to be parametrically  large).
The occurrence of the orientational moduli (dynamical fields of the $CP(N-1)$ model)
is a clear-cut indication of the non-Abelian nature of the string.
The tension of the above non-Abelian string is $T = 2\pi \xi$, up to small corrections.
This tension is lower than that of the Abrikosov string by $1/N$.

Classically, the $CP(N-1)$ model has a moduli field of vacua
and gapless excitations. The vacuum structure is drastically changed by 
nonperturbative infrared effects. As well-known, the vacuum of the model is unique,
the U(1)$\times$SU$(N-1)$ is restored, and the gapless excitations disappear.
From the four-dimensional point of view this means
that while geometrically the flux is oriented along the string,
fluctuations in the color space are gigantic.
The magnetic field inside the non-Abelian flux tube presented above
has no specific orientation in the SU$(N)$ space.

\section{What remains to be done}

Contours of the Big Picture are rather clear at present.  
Abelian confinement or its  absence in non-Abelian gauge theories  
 (in which the corresponding low-energy theory effectively becomes Abelian)
 seems possible to establish.  The instructive examples are provided by the
 Polyakov model with or without massless fermions, supersymmetric gauge theories, 
 and deformed QCD-like theories, including chiral theories.  In each case, 
 the detailed mechanisms explaining confinement are different;
 they are tied up, however, with certain topological excitations (see Table \ref{tab1}).  
 The Abelian  duality (either in three or four dimensions)  helps 
 us in each one of the above applications. 
Looking optimistically we can say that in a few years a much more crisp 
classification will arise along these lines.     

Still, there are dark corners.  We essentially do not know how the non-Abelian duality works. 
Dualizing non-Abelian Yang--Mills theories  with strings and linear confinement, and without any 
long-distance Abelian regime, is a challenge, regardless of whether the 
underlying theory is supersymmetric or not. 
A better understanding/description of the (presumably smooth) transition from
Abelian to non-Abelian confinement is badly needed. 
\begin{table}[htdp]
\begin{center}
\begin{tabular}{| p{4cm} |p{4cm}| p{4cm}|}
\hline
  & Seiberg--Witten & One-flavor QCD*   \\
\hline 
Control parameter & $ \mu / \Lambda  \equiv \epsilon$ & $L \Lambda \equiv \epsilon$ 
 \\ \hline 
$ \epsilon=0$  limit & no mass gap, no confinement & no mass gap, no confinement \\ \hline
(First) gauge symmetry breaking  & ${\rm SU}(N)\rightarrow {\rm U}(1)^{N-1}$ &   ${\rm SU}(N)\rightarrow {\rm U}(1)^{N-1}$ 
\\ \hline
$ \epsilon \ll  1$  regime & mass gap,  confinement &  mass gap,  confinement \\ \hline
Mechanism & Monopole condensation & Bion mechanism or instanton-monopoles  \\ \hline  
(Second) gauge symmetry breaking  & ${\rm U}(1)^{N-1} \rightarrow {\rm nothing}$ 
 & 
   ${\rm U}(1)^{N-1} \rightarrow {\rm nothing}$
   (or   discrete)
\\ \hline 
Expectation for $ \epsilon \gsim   1$  regime & non-Abelian  confinement &  
non-Abelian   confinement \\ \hline
\end{tabular}
\end{center}
\caption{ Main  features of the SW  and QCD* solutions ($N$ is not assumed to be large).   
}
\label{tab1}
\end{table}%

On a more practical side, it seems necessary to design a quasiclassical ``string theory,"
with  strings resembling those in the old Veneziano model, but modernized 
to somehow include  quarks at the endpoints
(it is highly desirable to have chiral symmetry of quarks built in,
perhaps by somehow adding pions separately), for the description of highly excited
mesons.
This string should be responsible not only for rotational excitations,
but for a mixture of rotational and oscillational (radial) excitations.
This is in addition to purely stringy excitations.
Then a description of highly excited mesons could be obtained.

\section*{Acknowledgments}

We are  grateful to A. Armoni and A. Yung for multiple and very useful
discussions.
M.S. is
supported in part
by DOE Grant DE-FG02-94ER-40823, and
 by {\em Chaire Internationalle de Recherche Blaise
Pascal} de l'Etat et de la R\'{e}goin d'Ile-de-France,
g\'{e}r\'{e}e par la Fondation de l'Ecole Normale Sup\'{e}rieure.
The work of  M.\"U. is supported by the U.S.\ Department of Energy Grant DE-AC02-76SF00515.

\end{document}